\documentclass[journal,letterpaper]{IEEEtran}

\usepackage{graphicx,cite,epsfig,amssymb,amsmath,amsfonts,multicol,subfigure,mathtools,bm,mathrsfs,setspace}
\usepackage{multirow}

\newcommand{\tabincell}[2]{\begin{tabular}{@{}#1@{}}#2\end{tabular}}

\begin{document}
\title{\huge Orthogonal Cocktail BPSK:  Exceeding Shannon Capacity of QPSK Input}
\author{
	
	Bingli~Jiao

	\thanks{B. Jiao is with the Department of Electronics and Peking University-Princeton University Joint Laboratory of Advanced Communications Research, Peking University, Beijing 100871, China (email: jiaobl@pku.edu.cn).}

}

\maketitle

\begin{abstract}

Shannon channel capacity of an additive white Gaussian noise channel is the highest reliable transmission bit rate (RTBR) with arbitrary small error probability.  However, the authors find that the concept is correct only when the channel input and output is treated as a single signal-stream.  Hence, this work reveals a possibility for increasing the RTBR further by transmitting two independent signal-streams in parallel manner.  The gain is obtained by separating the two signals at the receiver without any inter-steam interference.  For doing so, we borrow the QPSK constellation to layer the two independent signals and create the partial decoding method to work with the signal separation from Hamming to Euclidean space.  The theoretical derivations prove that the proposed method can exceed the conventional QPSK in terms of RTBRs.

\end{abstract}

\begin{IEEEkeywords}
reliable transmission bit rate, channel capacity, mutual information. 
\end{IEEEkeywords}

\IEEEpeerreviewmaketitle

\section{Introduction}
Shannon channel capacity underlies the communication principle for achieving the highest bit rate at which information can be transmitted with arbitrary small probability.  A standard way to model the input and output relations is the memoryless additive white Gaussian noise (AWGN) channel
\begin{eqnarray}
\begin{array}{l}\label{equ-ch}
y = x + n,
\end{array}
\end{eqnarray}
where $y$ is the received signal, $x$ is the transmitted signal and $n$ is the received AWGN component from a normally distributed ensemble of power $\sigma_N^2$ denoted by $n \sim \mathcal{N}(0,\sigma_N^2)$ \cite{Shannon1948}   .

In the previous literatures, the channel capacities of the finite alphabet inputs have been calculated in terms of the reliable transmission bit rates (RTBRs) by 
\begin{equation}
\begin{array}{l}\label{equ2}
\rm{I}(X;Y) =\rm{H}(Y) - \rm{H}(N)
\end{array}
\end{equation}
where $\rm{I}(X;Y)$ is the mutual information, ${\rm{H}}(Y)$ is the entropy of the received signal and ${\rm{H}}(N) = {\log _2} (\sqrt{2 \pi e \sigma_N^2})$ is that of the AWGN.   The some numerical results of \eqref{equ2} have calculated as shown in Fig. 1, where the capacity of Gaussian type signal input is also plotted as a reference.

\begin{figure}[htb]
	\centering
	\includegraphics[width=0.5\textwidth]{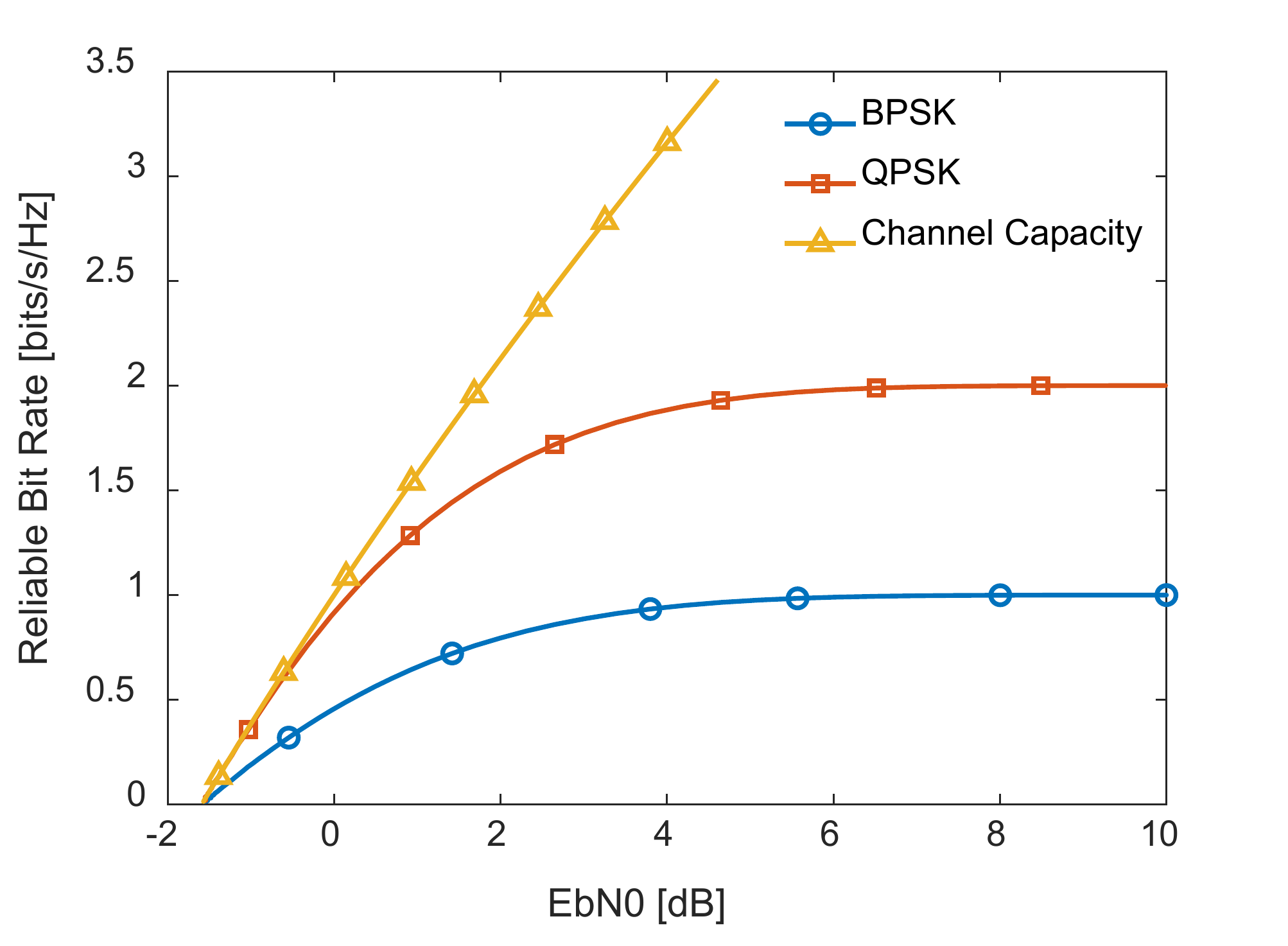}
	\caption{Reliable transmission bit rates for BPSK, QPSK and Gaussian type.}
	\label{fig1}
\end{figure}

Though the capacity concept holds for the last decades, there were still some
considerations on the possibility of beyond the capacities\cite{ jiao}.  A mathematical incentive can be found from the down-concavity of the mutual information curves as shown in Fig.1, from which one can conclude    
\begin{equation}
\begin{array}{l}\label{equ3}
\tilde{\rm{I}}_{x}\left[(E_1+E_2)/\sigma_N^2\right] < \tilde{\rm{I}}_{x_1}(E_1/\sigma_N^2)+\tilde{\rm{I}}_{x_2}(E_2/\sigma_N^2)  
\end{array}
\end{equation}
when $x=x_1+x_2$ is the signal superposition, $x_1$ and $x_2$ are two independent signals, and $E$, $E_1$ and $E_2$ are the symbol energies of $x$, $x_1$ and $x_2$, respectively.  In contrast to the conventional signal superposition methods, obtaining a gain from \eqref{equ3} requires non inter-symbol interference, i.e, non interference between $x_1$ and $x_2$.       

Nevertheless, the great difficulty can be encountered when one tries to organize the signal superposition that allows the separation to extract a contribution from \eqref{equ3}. 

This paper peruses, however, the inequality \eqref{equ3} by creating a new method, referred to as the orthogonal cocktail BPSK, that works in Hamming- and Euclidean space for separating the parallel transmission of the independent signals.  The derivations are done with the assumption of using the ideal channel codes that allow the error free transmission of BPSK and QPSK as well.  

Throughout the present paper, we use the capital letter to express a vector and the small letter to indicate its component, e.g.,  $A = \{a_1,a_2, ...., a_M\}$, where $A$ represents the vector and $a_i$ the $ith$ component.  In addition, we use $\hat{y}$ to express the estimate of $y$ at the receiver and $\tilde{\rm{I}}(\gamma)$ to express the nutual information $\rm{I}(X;Y)$ with SNR, $\gamma$, as the argument \cite{Verdu2007}.  The details are introduced in the following sections.

\section{Signal Superposition- and Separation Scheme}
Let us consider a binary information source bit sequence
which is partitioned into two independent subsequences expressed
in vector form of $C^{(i)} = \{c^{(i)}_1,c^{(i)}_2,.....,c^{(i)}_{K_i}\}$, where  $K_i$ is the length of the source subsequence and $i=1,2$ indicates the two source subsequences.

The two source subsequences are separately encoded, in Hamming space, by two difference channel code matrices 
\begin{eqnarray}
\begin{array}{l}\label{equ-code}
v^{(i)}_{m} = \sum\limits_{k_i}g^{(i)}_{mk_i}c^{(i)}_{k_i}  
\end{array}
\end{eqnarray}  
where $v^{(i)}_m$ is the $mth$ component of the channel code $V^{(i)}$, and $g^{(i)}_{mk_i}$ is the element of the code matrix $G^{(i)}$ for $i=1,2$ and $m=1,2,....,M$, respectively.  We note that $M$ is the length of the channel code word, and $R_1=K_1/M$ and $R_2=K_2/M$ are the two code rates which are unnecessarily to be equal.

For the signal modulations, we borrow the QPSK constellation to map the two channel codes, $V^{(1)}$ and $V^{(2)}$, into the Euclidean space specified by  
$s^{(1)}=\{\sqrt{2}\alpha,j0\}$, $s^{(2)}=\{0, j\sqrt{2}\alpha\}$, $s^{(3)}=\{-\sqrt{2}\alpha,j0\}$ and $s^{(4)}=\{0, -j\sqrt{2}\alpha\}$, where $\alpha>0$ and $j=\sqrt{-1}$ , as shown in Fig.2.  

In contrast the conventional QPSK modulation, the proposed method allows $V^{(1)}$ to be demodulated and decoded separately from $V^{(2)}$.  This decoding scheme is defined as the partial decoding in this approach because that only one source subsequence, i.e. $C^{(1)}$, is decoded.  

Consequently, using the decoding results of $C^{(1)}$ allows a reliable separation of $V^{(2)}$ from $V^{(1)}$.  Then, $V^{(2)}$ can be demodulated over two perpendicular BPSKs: one is constructed by $s^{(1)}$ and $s^{(3)}$ and the other by $s^{(2)}$ and $s^{(4)}$.  

More important, the Euclidean distance between the two signal points with each BPSK from the decouple is larger than $2\alpha$ that results, eventually, in a RTBR gain as found latter.  Thus, we refer the proposed method to as the orthogonal cocktail BPSK (OCB), as explained in the following paragraphs.       

\begin{figure}[htb]
	\centering
	\includegraphics[width=0.3\textwidth]{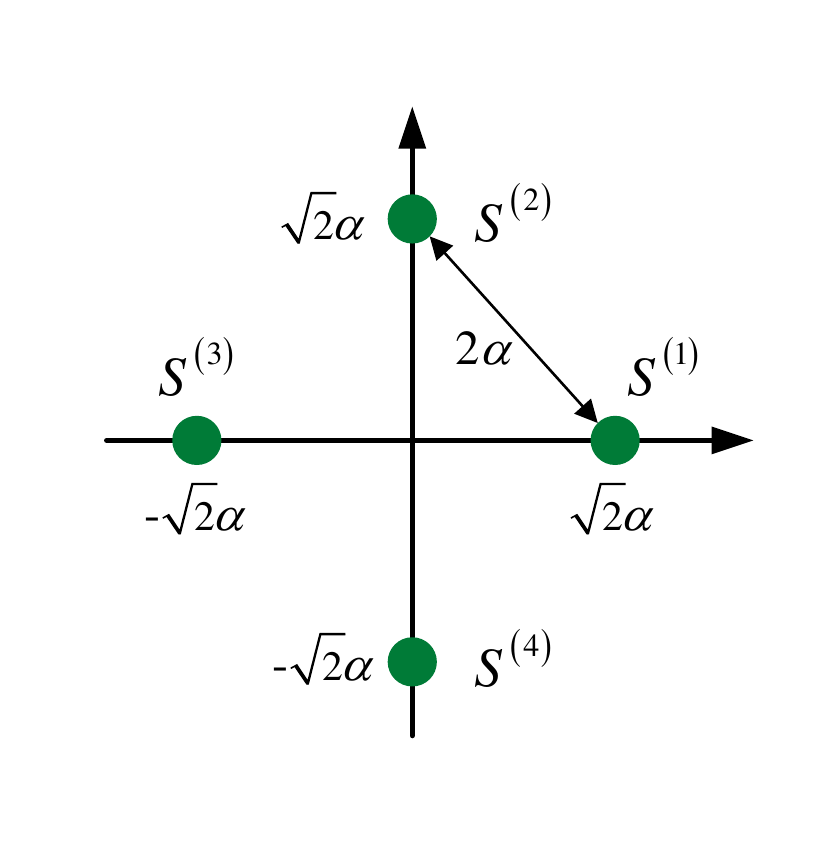}
	\caption{Constellation of the proposed method.}
	\label{fig2}
\end{figure}

The OCB modulation is classified into two cases with respect to the bit values of $v^{(1)}_m =0$ or $1$.  Case I belongs to $v^{(1)}_m=0$, whereby we map $v^{(1)}_m=0$ and $v^{(2)}_m=0$ onto $s^{(1)}_m$, and $v^{(1)}_m=0$ and $v^{(2)}_m=1$ onto $s^{(3)}_m$.  Actually, one can regard that the BPSK in horizontal direction is used to the signal mapping of case I. 

Case II belongs $v^{(1)}_m=1$, whereby $v^{(1)}_m=1$ and $v^{(2)}_m=0$ are mapped onto $s^{(2)}_m$, and $v^{(1)}_m=1$ and $v^{(2)}_m=1$ onto $s^{(4)}_m$, where one can find the BPSK in vertical direction. 

For expressing the OCB modulations more intuitive, the signal mapping of the two cases is listed in Table I, in which $s^{(\kappa)}_m$ for $\kappa=1,2,3, 4$ is the QPSK constellation with sequential index $m$ added.

\begin{table}[!t]
	\renewcommand{\arraystretch}{1.5}
	\centering
	\small
	\caption{Signal modulation results.}
	\label{Table1}
	\begin{tabular}{c|c|c|c}
		\hline
		\multirow{2}{*}{Case I} 
		& \tabincell{c}{$v^{(1)}_m=0$} & \tabincell{c}{$v^{(2)}_m=0$} & \tabincell{c}{$S^{(1)}_m$} \\
	    \cline{2-4}
		& \tabincell{c}{$v^{(1)}_m=0$} & \tabincell{c}{$v^{(2)}_m=1$} & \tabincell{c}{$S^{(3)}_m$} \\
		\hline
		\multirow{2}{*}{Case II} 
		& \tabincell{c}{$v^{(1)}_m=1$} & \tabincell{c}{$v^{(2)}_m=0$} & \tabincell{c}{$S^{(2)}_m$} \\
	    \cline{2-4}
		& \tabincell{c}{$v^{(1)}_m=1$} & \tabincell{c}{$v^{(2)}_m=1$} & \tabincell{c}{$S^{(4)}_m$} \\	
		\hline
		
	\end{tabular}
\end{table}

Then, the transmitter inputs one symbol another into the AWGN channel by 
\begin{eqnarray}
\begin{array}{l}\label{equ11}
y_m = s^{(\kappa)}_m + n_m,    \  \  \ for \ \ m\ =\ 1, \ 2,\, ....,M
\end{array}
\end{eqnarray}
where $y_m$ is the received signal, and $s^{(\kappa)}_m$ the transmitted symbol and $n_m$ is the Gaussian noise statistically equivalent to that in \eqref{equ-ch}.

At the receiver, all received signals in Euclidean space are recoded sequentially.  The demodulation starts from $V^{(1)}$ by  
\begin{eqnarray}
\begin{array}{l}\label{equ-v1}
\hat{y}_m = s^{(1)}_m \ \ \ or \ \ s^{(3)}_m\ \ \ \  for \ \ \ v^{(1)}_m=0 
\end{array}
\end{eqnarray}
and 
\begin{eqnarray}
\begin{array}{l}\label{equ-c1}
\hat{y}_n = s^{(2)}_m \ \  or \ \ s^{(4)}_m\ \ \ \ for \ \ \ v^{(1)}_m=1  
\end{array}
\end{eqnarray}
where $ \hat{y}_m $ is the estimate of $ y_m $. Then,  we work on the partial decoding scheme defined above by using the estimates of \eqref{equ-v1} and \eqref{equ-c1} to obtain $\hat{C}^{(1)}$.

Once $\hat{C}^{(1)}$ has been obtained, the receiver reconstructs the channel code $V^{(1)}$ by 
\begin{eqnarray}
\begin{array}{l}\label{equ-re1}
\hat{v}^{(1)}_m =  \sum\limits_{k_i}g^{(1)}_{mk_1} \hat{c}^{(1)}_{k_1},
\end{array}
\end{eqnarray}
which can be used to decouple the QPSK into the two perpendicular BPSKs in Euclidean space.  

\begin{figure}[ht]
	\centering
	\subfigure[]{
		\includegraphics[width=0.3\textwidth]{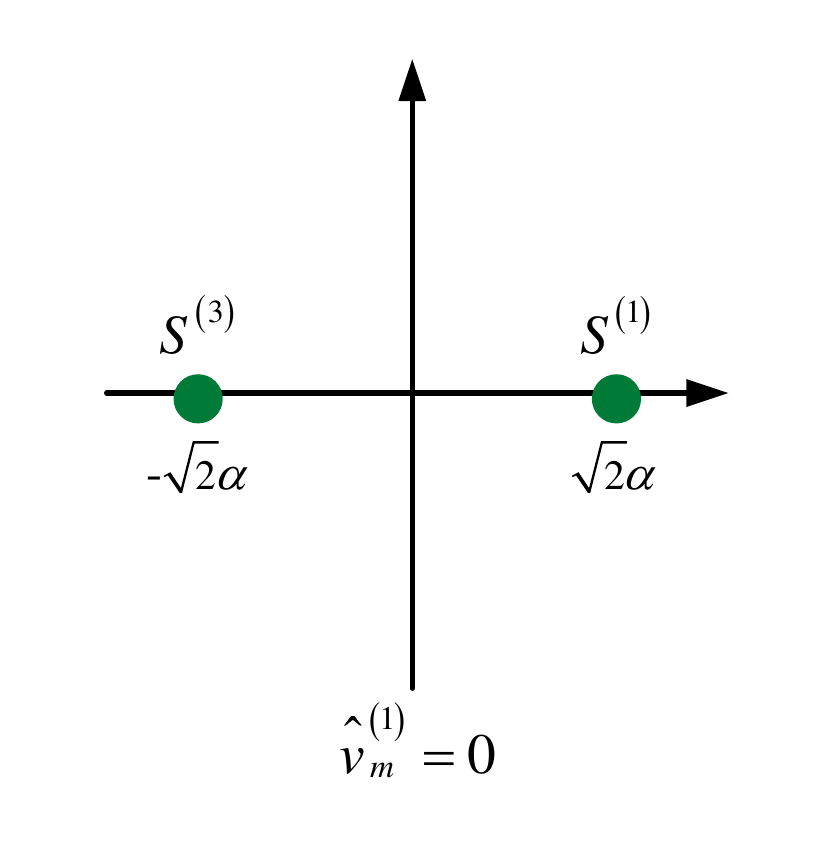}
		\label{fig3a}}
	\subfigure[]{
		\includegraphics[width=0.3\textwidth]{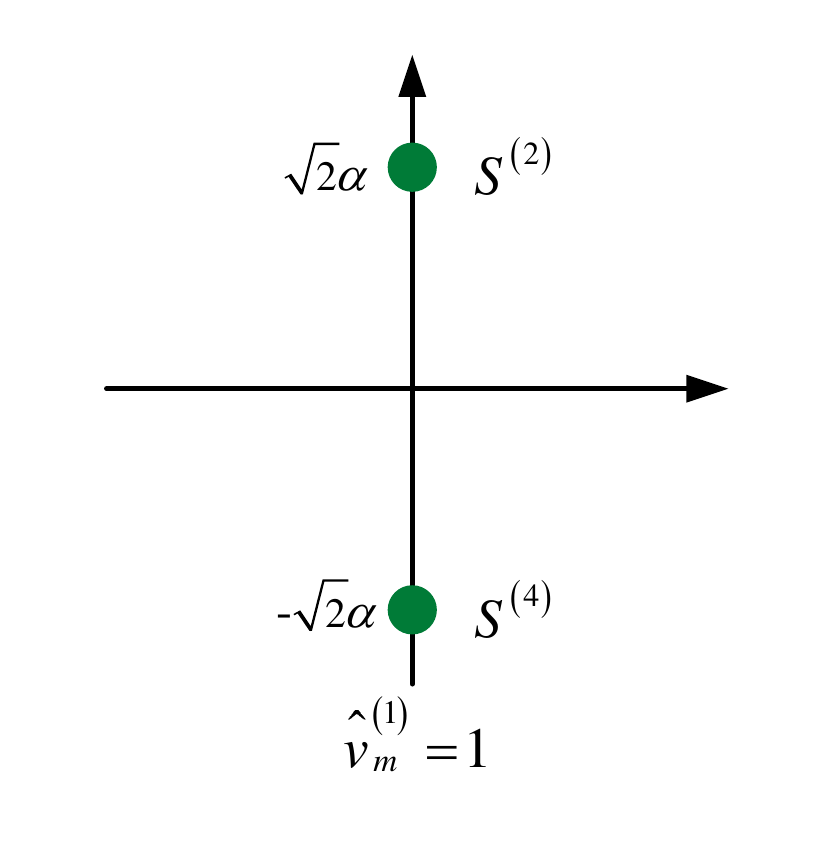}
		\label{fig3b}}
	\caption{The mapping symbols for (a) case I and (b) case II.}
	\label{fig3}
\end{figure}

The results of \eqref{equ-re1} can be regarded as the reliable reconstruction, whereat $\hat{v}^{(1)}_m = 0 $ indicates that the recoded signal belongs to case I, while $\hat{v}^{(1)}_m = 1$ to case II.  Thus, the two perpendicular BPSKs can be decoupled as shown in Fig.3(a)(b), receptively.   This allows the detection of $v^{(2)}_m$ as follows.
    
If $v^{(1)}_m=0$, the receiver detects the recoded signal $s^{(\kappa)}_m$ by  
\begin{eqnarray}
\begin{array}{l}\label{equ-v21}
\hat{y}_m = s^{(1)}_m \ \  for\  \  v^{(2)}_m=0 
\end{array}
\end{eqnarray}
and 
\begin{eqnarray}
\begin{array}{l}\label{equ-v2}
\hat{y}_m = s^{(3)}_m\ \  for\  \  v^{(2)}_m=1 
\end{array}
\end{eqnarray}
If $v^{(1)}_m=1$, the receiver detects $v^{(2)}_m$ by  
\begin{eqnarray}
\begin{array}{l}\label{equ-v31}
j\hat{y}_m = js^{(2)}_m \ \  for\  \  v^{(2)}_m=0 
\end{array}
\end{eqnarray}
and 
\begin{eqnarray}
\begin{array}{l}\label{equ-v3}
j\hat{y}_m = js^{(4)}_m\ \  for\  \  v^{(2)}_m=1 
\end{array}
\end{eqnarray}
Then, by taking the estimates of \eqref{equ-v21}, \eqref{equ-v2}, \eqref{equ-v31} and \eqref{equ-v3} to the decoding of $C^{(2)}$, we can obtained the  $\hat{C}^{(2)}$.

In practical situation, when an error presents in reconstruction of $\hat{v}^{(1)}_m$, the detection of $v^{(2)}_m$ can be wrong with $50\%$ probability.   Then, the decoding can suffer from the error propagation. 

However, when working with the ideal low density block code, the infinitive error probability of $\hat{C}^{(1)}$ can lead to the infinitive small probability of the reconstruction of $\hat{V}^{(1)}$.  Thus, the error rate problem in the signal separation can be neglected when we are studying on the capacity issue.

\section{Up-Bound Issue}
Assume that we are working with the ideal channel codes that allows error free transmissions of QPSK and BPSK, the RTBR of the OCB method is found higher than that of the QPSK input as proved in the following paragraphs.     

First, we prove that the RTBR of $C^{(1)}$ is at a half of QPSK input by    
\begin{eqnarray}
\begin{array}{l}\label{equ12}
\mathbb{R}_{c1} = \frac{1}{2}\tilde{\rm{I}}_q(2\alpha^2/\sigma_N^2)
\end{array}
\end{eqnarray}
where $\mathbb{R}_{c1}$ is the RTBR of $C^{(1)}$ and $\tilde{\rm{I}}_q$ is the mutual information of QPSK input.  

Proof: 
In order to prove this issue, we first recall the following theorem: when the Euclidean distance $\tilde{d}(\xi,\xi')$ is the same, the large Hamming distance of the channel codes can lead to smaller BER.  This is true when we compare the OCB with the conventional BPSK since the source codes, i.e, $C^{(1)}$, can be found as a QPSK coded modulation that deleted a half of the source bits.  The OCB can have the smaller BER in comparison with that of QPSK input.  Thus, for using infinitive long channel codes, whenever the transmission of QPSK input is of infinitive small error probability, the partial coding with $C^{(1)}$ applies as well.           

The RTBR of $C^{(1)}$ is at half of the QPSK input because the former transmits one channel bit per symbol, while the latter two channel bits.  

Once $C^{(1)}$ is transmitted to the receiver without error, the demodulation of $V^{(2)}$ can be done by using the two BPSK symbols, in each of which the Euclidean distance is $\sqrt{2}\alpha^2$.  Thus, the symbol energy is found at $2\alpha^2$ that should be used to calculate the mutual information       
\begin{eqnarray}
\begin{array}{l}\label{equ15}
\mathbb{R}_{c2}=\tilde{\rm{I}}_b(2\alpha^2/\sigma_N^2)
\end{array}
\end{eqnarray}
where $\mathbb{R}_{c2}$ is the RTBR of $C^{(2)}$, and $\tilde{\rm{I}}_b$ is the mutual information of BPSK.

Finally, the summation of RTBRs in \eqref{equ12} and \eqref{equ15} yields
\begin{eqnarray}
\begin{array}{l}\label{result2}
\mathbb{R}_J = \frac{1}{2}\tilde{\rm{I}}_q(2\alpha^2/\sigma_N^2)+\tilde{\rm{I}}_b(2\alpha^2/\sigma_N^2)
\end{array}
\end{eqnarray}
where $\mathbb{R}_J$ is the RTBR of this approach.  

The numerical results of \eqref{result2} are plotted in Fig.4, whereat one can  that the curve of OCB on the left side of QPSK. This indicates the RTBR exceeding of QPSK input.    

\begin{figure}[htb]
	\centering
	\includegraphics[width=0.5\textwidth]{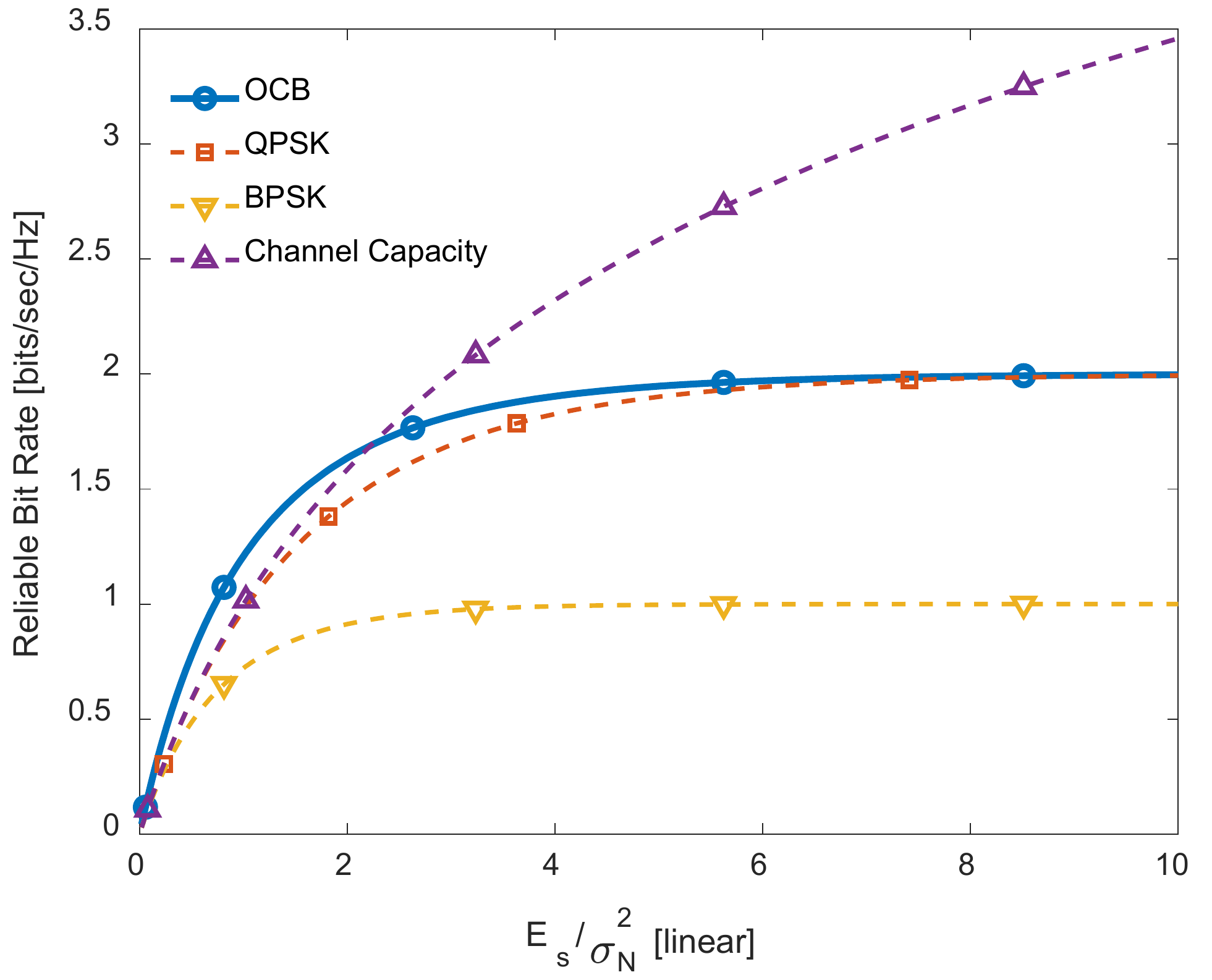}
	\caption{ADRs of OCB compared with QPSK and BPSK versus linear ratio of $E_s/\sigma_N^2$.}
	\label{fig4}
\end{figure}
  
\section{Conclusion}

In this paper, we proposed the OCB method for increasing the RTBR further beyond the QPSK input and, even, the Shannon capacity of Gaussian type signals.  The proposed method works in Hamming and Euclidean space in separation of the two independent signals transmitted in parallel over an AWGN channel. Theoretical derivations prove this approach base on the assumption of using the ideal channel codes.

\end{document}